\def\sun{\hbox{$\odot$}}

\documentclass[cup5b]{caps}
\usepackage{graphicx}
\usepackage{amssymb}
\usepackage{aaas_head}  %Use this for contributed papers.

\HeadText{J. Schombert}  %Enter author name; if three authors, list all three;
\begin{document}

\pagenumbering{arabic}

\author[]{J. SCHOMBERT\\
Dept. of Physics, University of Oregon}

\chapter{The Evolution of Galaxies: A Metaphysics Viewpoint}

\begin{abstract}

This is the written version of an invited review talk for the 13 Feb 2004 AAAS Meeting in
Seattle. The talk's goal is to present a philosophical view of extragalactic astronomy as it
applies to the sub-field of galaxy evolution.  The talk is divided into three parts: 1) How we
got to where we are (technology drivers to our science goals), 2) What's new and special (how
that technology has achieved our recent science results) and 3) How an improved worldview will
help us in the near future.  The intended audience for this talk is a generally knowledgeable
scientist, but not an astronomer by training.  This talk is also *not* intended to be a
complete review of the field of galaxy evolution and only includes a few recent results
extracted from the astro-ph archives to present the current state of our field.

\end{abstract}

\section{Technology and the Golden Age of Galaxy Evolution}

Historians will no doubt look back onto the last part of the 20th century as a true "Golden
Age" for observational cosmology and galaxy evolution due to the enormous advancements driven
primarily by technological leaps in the telescopes, detectors and computers.  Like most modern
sciences, the study of galaxies is beyond the basic human senses (although the Magellenic
Clouds are visible to the naked eye in the southern hemisphere and Andromeda is at the limit
of the human eye for very dark sites).  Thus, the study of galaxies did not even begin until
the development of 1m class telescopes at the end of the 19th century.  With the discovery of
an expanding Universe (Hubble 1936) and the confirmation of a Creation point (Penzias \& Wilson
1965), it was quickly realized that astronomers have a unique capability, that no other science
has. This is the capability to actually follow the evolution of distant objects while the size
of the Universe is much greater than the speed of light (i.e. lookback time).  To imagine the
impact of lookback time on a field of science, imagine the response of a palaeontologist if
offered a device that would allow them to observe animal behavior during the Jurassic era, or
a musicologist who is allowed to listen in on Mozart's practice sessions.

Lookback time means that the more distant an object, the farther into our past it is. Thus
began in extragalactic astronomy the great hunt for high redshift objects, redshift being a
measure of distance in an expanding Universe.  Of course, more distant also means fainter in
apparent luminosity, which drives the need for larger and larger telescopes to collect the
photons from high redshift, or very distant, galaxies.  Our first quantum jump in the study of
galaxy evolution was during the 1950's and 60's with the development of the Palomar 5m and
NOAO's 4m telescopes.  With this technology, spectroscopy out to several billion light-years
was doable.  Solid state technology in the 80's gave us highly efficient and highly reliable
(for photometry) detectors.  The 90's gave us 10m class telescopes and the Hubble Space
Telescope, arguably the most important instrument for the study of galaxy evolution.

Parallel to observational achievements were breakthroughs in many theoretical arenas, primarily
the coming of age of the computer simulation in the 80's as the main tool for astrophysical
research (observers call themselves astronomers, while theorists call themselves
astrophysicists for some strange reason).  During the 90's, decreasing access to telescopes
(due to an increasing population of observers) and decreasing computer hardware costs produced
a glut of theoretical Ph.D. theses.  Within a decade, our field went from the point where
theorists numbered one for every four observers to near parity.  When combined with uncertain
values for the basic cosmological constants (i.e. $H_o$ and $q_o$), this lead to an explosion
of ideas, but with very few constraints to narrow our focus on what was really happening after
galaxy formation.

Unfortunately, the same technology that allows us to observe the past also confines our window
of knowledge.  For example, galaxies present themselves over a range of luminosities,
which corresponds directly with the mass of the galaxy (i.e. the number of stars).  While the
brightest galaxies (up to 10$^{13}$ $M_{\sun}$) are the most dominant members of clusters of
galaxies, and the easiest to obtain photometric and spectroscopic information, they are not
the dominant members of the Universe by number.  That honor goes to the small, dwarf galaxies
(masses around 10$^8$ $M_{\sun}$) who number over a factor of 10 to 100 more numerous than the
brightest galaxies in clusters.  Thus, the limiting magnitude of your telescope (the depth in
luminosity that can be achieved for a decent S/N) will allow you to study bright galaxies to
distant redshifts (far into the past) or both bright and faint galaxies at lower redshifts.
This bias forced our conclusions about galaxy evolution in the 80's and 90's to be limited to
only the very brightest galaxies in rich environments.  The newest technology allows us to
address the full galaxy population at cosmologically interesting distances of 0.5 to 0.7 (see
Figure 1.1).

\begin{figure}
\centering
\includegraphics[width=12cm]{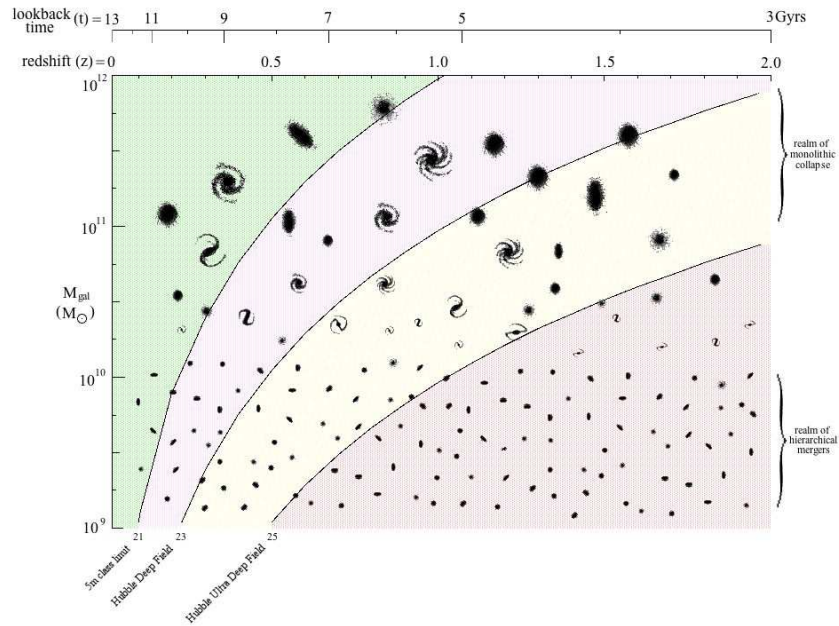}
\caption{
This figure displays the effects of limiting magnitude to the ability of galaxy evolution
studies to resolve their science goals.  Early technology (shown by the 5m limit) shows that
only the brightest galaxies could be measured at cosmologically interesting distances.  The
Hubble Deep Field (HDF) extended the depth to more normal galaxies and the Hubble Ultra Deep
Field (HUDF) will allow a comparison of the evolution of dwarf and giant galaxies, a key test
to the hierarchical merger scenario of galaxy formation.
}
\end{figure}

The greatest progress with respect to the study of galaxy evolution was the determination
of the cosmic distance scale parameters, $H_o$ and $q_o$ by the Hubble Cepheid Key Project and
the High Redshift Supernova Cosmology Project.  While these two parameters are not directly
related to galaxy evolution parameters, their determination has finally released researchers
from being forced to interpret evolutionary effects through a myriad of combinations of $H_o$
and $q_o$.  Or, worse, attempting to deduce these cosmological parameters from poorly
known evolutionary effects.  Galaxy evolutionists are now free to focus solely on the star
formation history of galaxies with lookback time determined solely by the redshift of the
galaxy.

\section{Current State of Galaxy Evolution}

While lookback time allows the study of galaxy evolution, there was little hope in the early
1980's that our technology would allow us to take advantage of our unique Universe.  The
simplest evolutionary models in the 70's, plus the ages of globular clusters, indicated that
galaxies formed very soon after the Big Bang and their star formation was completed soon
afterwards in order to account for their current photometric properties.  Thus, evolutionary
effects would not be seen at intermediate redshifts ($z = 0.2$ to $0.8$) but only at redshifts
greater than 1.  The 4 to 5m class telescopes at the time, with early CCD technology, had
insufficient limited magnitude to investigate this realm (again see Figure 1.1).

The field of galaxy evolution received a huge boost with the discovery by Butcher \& Oemler in
the early 80's of large numbers of blue galaxies in clusters with relatively low redshifts ($z
\approx 0.3$).  When compared with typical galaxies in present-day clusters, the rapid change
in color for the Butcher-Oemler population is unexpected and the identification of the
progenitors an important problem.  Blue colors in galaxies can arise through numerous
processes; massive star formation, non-thermal emission (AGN) or a skewed metal-poor stellar
contribution, but spectroscopy quickly confirmed that the Butcher-Oemler population was due to
star formation effects.

While the degree of strength of the Butcher-Oemler effect is still debated (for example, the
fraction of blue galaxies with redshift varys sharply depending on bandpass of observation),
it is well established that there exist significant numbers of blue galaxies in clusters at
redshifts of 0.2 to 0.3.  In contrast, blue galaxies are extremely rare in rich clusters at
the present epoch.  This is the first instance of a violation of the strong form of the
Copernican Principle in astronomy.  In its strong form, the Copernican Principle is a valuable
philosophical tool to guide our interpretation of data.  It is simply stated that if a theory
requires a special origin or viewpoint, then it is not plausible or, at least, its plausibility
is in question.  The classic historical example is, of course, the falsification of the
geocentric model of the solar system.  However, the Copernican Principle has been applied many
times in the history of science.  For example, the theory of evolution removed humankind from
its special place in the hierarchy of lifeforms.

With respect to Butcher-Oemler effect, the discovery of new population of galaxies in the recent
past makes the current population of galaxies in clusters appear to be special.  The spatial
version of the Copernican Principle is still in effect, the Universe appears to look the same
at any particular epoch (i.e. the cosmological principle).  But, the Butcher-Oemler effect
violates the temporal component of the Copernican Principle, which is not too surprising since
the Universe does evolve with time.  Evolution, with respect to star formation in galaxies, is
a finite process since there is only a finite supply of gas in galaxies.  And it is clear that
the Butcher-Oemler galaxies have not disappeared, they have simply become faded LSB dwarfs or
converted into S0 galaxies.  So it is just a coincidence that we live in the epoch were cluster
disk galaxies have exhausted their gas supplies.

The discovery of rapid evolution in rich clusters divided the field of galaxy evolution into
studies of the `blue' population in clusters, those star-forming or active systems, versus the
`red' population, galaxies which, supposedly, formed all their stars in an early, massive burst
and have remained quiescent to the present epoch.  Both populations had their unique
characteristics to help address various problems in galaxy evolution.  For example, the blue
population was well positioned to investigate the effects of environment on the star formation
history of a galaxy.  The red population would provide a window into the epoch of galaxy
formation by tracing the color evolution of ellipticals guided by spectroevolution models.

Deep photometric studies in the 80's and 90's demonstrated that the Butcher-Oemler effect
increased with more distant clusters, but not in a strictly linear manner where various
environmental effects have a significant, and random, effects (astronomers also have our own
nature versus nurture debate).  Ram pressure stripping by the hot intracluster x-ray gas and
cluster tidal effects (so-called galaxy harassment) both play roles of disrupting and exciting
the star formation process in galaxies.  Later imaging studies indicated that the blue
population was divided into bright, massive, star-forming galaxies and a fainter, dwarf
starburst population which explains the sensitivity to environmental effects is due, in part,
to the range of masses involved in the Butcher-Oemler effect.

The red population, presumingly the ancestors of the passive, red ellipticals found in the
local Universe, originally served as tracers of the epoch of galaxy formation.  All indications
from photometric and spectroscopic observations of nearby ellipticals is that they harbor the
oldest stellar populations and, therefore, following their color evolution is the simplest
route to the epoch of initial star formation of all galaxies.  Their spectrophotometric
properties are well matched to our expectations from evolution models of a single age, single
burst formation origin (the so-called monolithic collapse scenario).  However, these results
are very difficult to interpret due to the bias manner in which we select our red
population (i.e. we select against blue objects, meaning younger).  Increased depth in
luminosity allows us to follow the color-magnitude relation for ellipticals (the correlation
between the mass of a galaxy, the depth of its gravitational well, and its ability to retain
enriched gas from SN explosions, the mean chemical composition of the underlying stellar
population) which is a less biased way to observe the red population as a whole.

While larger telescopes have increased the depth and accuracy of our spectrophotometric color
studies, the two most dramatic advancements to galaxy evolution studies has been 1) the ability
to follow galaxy morphology to high redshifts (improved depth and resolution of HST) and 2) the
ability to measure the dynamical properties of distant galaxies (through the increased
collecting power of Keck and Gemini).

\begin{figure}
\centering
\includegraphics[width=12cm]{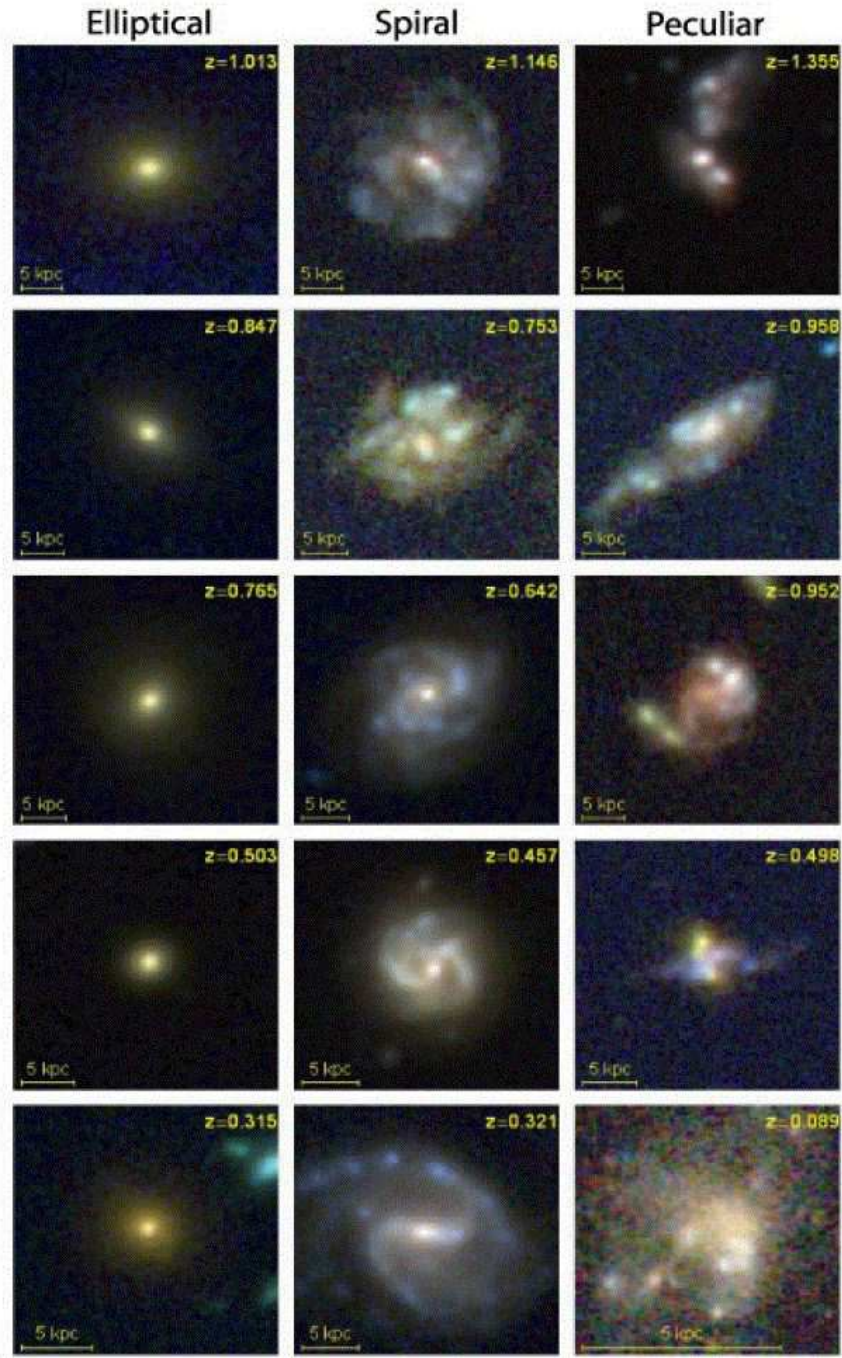}
\caption{
A visual comparison of galaxy morphological evolution with redshift (taken from Abraham and van
den Bergh 2001).  Ellipticals, spirals and irregulars at various redshifts taken from the
Hubble Deep Field.  Nearby galaxies are at the bottom with increasing distance (therefore,
lookback time) towards the top.  While ellipticals remain unchanged with time, spirals lose
pattern organization and increase in fragmentation of spiral arms.
}
\end{figure}

Galaxy morphology is typically considered an archaic field, where the shape and pattern of a
galaxy are used to provide a classification scheme (spiral, elliptical, irregular).  However,
there are clear and distinct correlations between morphological appearance and star formation
history.  For example, galaxies undergoing current star formation have spiral or irregular
shapes and appearances.  Old, quiescent galaxies are smooth and elliptical in appearance.  In
the field of galaxy evolution, the morphological appearance of distant galaxies was a question
placed aside in the past due to the lack of spatial information.  Distant galaxies were
unresolved at redshifts greater than 0.2 from ground-based telescopes.  This changed,
dramatically, with the repair of HST in the early 90's and morphological evolution has taken
its place along side color evolution as a key tool in understanding the star formation history
of galaxies.  In addition, new and improved software has matured our
image analysis tools.  Image analysis now goes hand-in-hand with galaxy classification with the
introduction of quantifiable parameters such as concentration, asymmetry and clumpiness
indices.

A visual summary of the morphological evolution can be seen in Figure 1.2 (taken from Abraham
\& van den Bergh astro-ph/0203042).  While smooth, red ellipticals are found at all epochs in
the past, the grand design spirals appear to be deficient and/or missing.  Again, this seems to
be a violation of the Copernican principle (that we live in a `special' time of grand design
spirals), in fact this just represents the time for spiral wave patterns to develop.  It is no
particular surprise to find fewer well-developed spiral density waves in the past any more than
it is surprising to find the current population of Sa's to be running low on neutral hydrogen
gas (the source of star formation) as they will soon transition into smaller bulge S0's.

Combining color and morphological studies can result in information such as that presented in
Figure 1.3, a spatial plot of color and morphological type for a rich cluster at a redshift of
0.3.  In this figure, the contrast between expectations based on morphology can be compared to
the actual stellar population that makes up the galaxy.  For example, most blue ellipticals are
found at the cluster edges.  And what few red spirals there are, are located in the cluster
core.  Clearly the cluster environment has an inhibiting effect on global star formation in a
galaxy, or the entrance into a cluster environment sparks a surge in star formation.

\begin{figure}
\centering
\includegraphics[width=12cm]{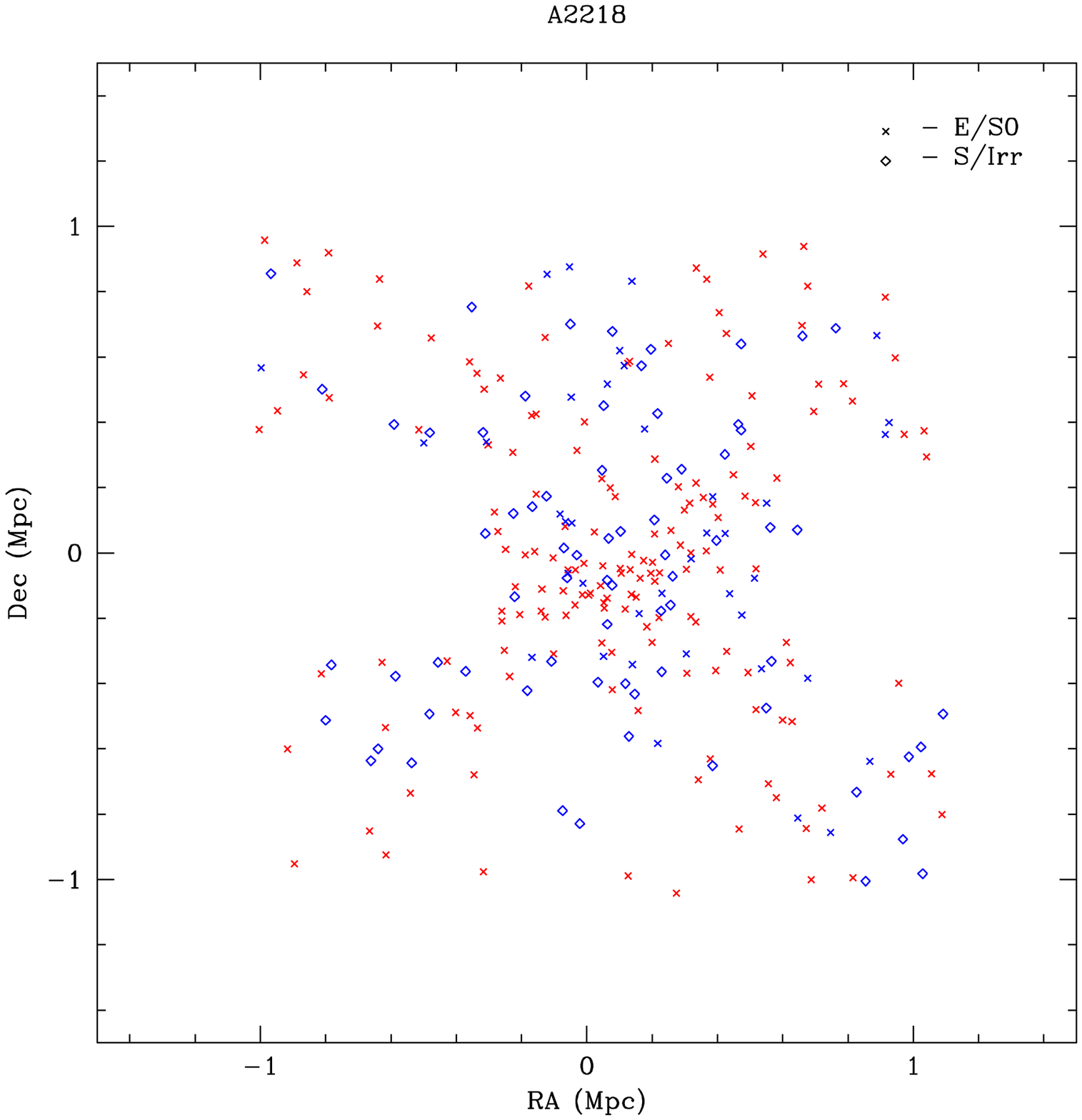}
\caption{
A example of the combination of spectrophotometric and morphological information for a rich
clusters at a redshifts of 0.2.  The morphological classification is given as the symbol shape,
the color classification is given as the symbol color.  While elliptical morphologies can be
found throughout the cluster, the red ellipticals are concentrated in the core and the blue
ellipticals are found primarily around the cluster edges.  Likewise, blue galaxies in general
avoid the core, demonstrating the hostile environment for star-forming galaxies and the effect
of encountering the cluster environment for infalling systems.
}
\end{figure}

Perhaps the greatest achievement of our new technology telescopes is the acquisition of
dynamical information for distant galaxies up redshifts of 1.2.  Information regarding the
rotation curve of disk galaxies, or the velocity dispersion of ellipticals, is the `holy grail'
of galaxy studies as it provides information on the mass of the galaxy, and thereby the
mass-to-light ratio and the amount of dark matter.  Dynamical and structural studies of
galaxies are summarized by the information presented in the cleverly named Fundamental Plane.
Every galaxy formation scenario must provide an explanation for the FP, its correlation between
galaxy properties and its zero-point values.  Thus, the ability to follow the evolution of the
FP brings galaxy evolution work in alignment with our knowledge of nearby galaxies, completing
the circle.

\section{A Philosophy for Galaxy Evolution}

In this last section I will argue that the field of galaxy evolution has lost some of its
connection to the traditional method of science and there is some concern that non-scientific
pressures are degrading the quality of our projects.  First off, it has been argued that
astronomy is, in fact, not a science as given by the rigid definition of science in terms of
experimentation and repeatability.  Experimentation is considered to be the core of hypothesis
testing and building a framework for understanding any physical phenomenon is accomplished by
hypothesis testing.  While it is true that there is very little to astronomy that can be
investigated by experimentation (an H-bomb is the closest we come to making a star), astronomy
is still a science by virtue of the logical techniques that we apply (it must be since it
counts as science credit for humanities majors).

We are certainty a science by Baconian standards as our main focus is the search for
regularities based on facts. And, where experimentation is the foundation of physics,
observation is the foundation of astronomy.  A lack of experimentation strengthens the
importance of repeatability for astronomy, where repeatability becomes verification and the
scientific process is one of deducing general properties from numerous, multi-wavelength
observations.

While observations are of critical importance, astronomy is not without a solid theoretical
framework.  For example, a well developed theory of stellar evolution has progressed into
stellar population models in which we can compare and interpret our galaxy evolution
observations.  Observations are interpretations of facts observed, interpreted through our
models and scenarios.  Our need for theory is critical, for observations cannot proceed theory
as some idea must be presupposed by any observation.  There are frequent instances where some
discovery existed in the data but was missed because we lacked a foundation to even identify
its existence.  Likewise, our theoretical community has a great deal of freedom to come up with
and express new ideas and theories, but not free in the same sense as art or literature, for
these ideas must survive confrontation with observation and it is this interplay between theory
and observation that makes astronomy a robust physical science.

However, due to the asymmetry in astronomical research (only passive observations), we, as a
field of study, have focused on a Popperian idea of science with an emphasis on testability
and falsification over our ideas/models/scenarios.  Astronomy is not a collection of theories
and models which are true, but rather a system of ideas that are `best'.  Where best is defined
as closer and closer to the truth (although we may never know if we have reached our goal).
Thus, the more testable an idea is, the better it is for the field.  While astronomical
observations are passive, their strength lies in their testability and access, i.e. they are
available for refutation by anyone.  More access means more exploration of the data which leads
to new discoveries and new ideas.

Access and funding issues have severely, perhaps fatally, interfered with this process.
Access to oversubscribed telescope time has produced a situation where the typical
observer can only obtain a few nights per year or a handful of orbits on HST.  Funding faces
the same difficulty with all NSF and NASA programs oversubscribed to at least the 5-to-1 level.
However, there is some hope that astronomy as a field of research can regain the high ground
of scientific certainty through new access, so-called e-science.

This is a particularly critical time to reexamine our philosophical methods with the advent of
numerous digital sky surveys and the Virtual Observatory incentives.  Tools are available for
inspection and analysis of many datasets.  The data is out there waiting to be analyzed,
discoveries are already on hard disk.  And it should be considered a noble goal to re-examine
published ideas and even re-analyze objects with previous observations.  A trained and tool
savvy astronomer is the astronomer of the future, although we now live under the danger of
using `black box' reduction or analysis routines with unknown errors.  The power in the next
generation e-scientist is his/her abilities to understand the data and its limitations and
target the right tools to the right questions.  In fact, since the data is unlimited, the skill
will be in coming up with the right questions to match the answers the data can provide.

\end{document}